\documentclass[12pt]{iopart}

\usepackage{color}
\usepackage{graphicx}
\usepackage{url}
\usepackage{hyperref}	
\definecolor{greencolor}{rgb}{0,0.5,0.2}
\definecolor{redcolor}{rgb}{0,0.,0.}
\definecolor{bluecolor}{rgb}{0,0.,1.}
\definecolor{greycolor}{rgb}{.5,.5,.5}
\def\Red#1{{\color{redcolor} #1}}

\begin{document}

\title[Robustness of community structure to node removal]{Robustness of community structure to node removal}

\author{Diego R. Amancio}
\address{Institute of Mathematical and Computer Sciences\\
University of S\~{a}o Paulo \\
S\~{a}o Carlos, S\~ao Paulo, Brazil\\}
\ead{diego.raphael@gmail.com, diego@icmc.usp.br \\ }

\author{Osvaldo N. Oliveira Jr. and L. da F. Costa}
\address{S\~{a}o Carlos Institute of Physics\\
University of S\~{a}o Paulo \\
S\~{a}o Carlos, S\~ao Paulo, Brazil \\ \ \\}

\vspace{10pt}

\begin{abstract}
The identification of modular structures is essential for characterizing real networks formed by a mesoscopic level of organization where clusters contain nodes with a high internal degree of connectivity. Many methods have been developed to unveil community structures, but only a few studies have probed their suitability in incomplete networks. Here we assess the accuracy of community detection techniques in incomplete networks generated in sampling processes. We show that the walktrap and fast greedy algorithms are highly accurate for detecting the modular structure of incomplete complex networks even if many of their nodes are removed. Furthermore, we implemented an approach that improved the time performance of the walktrap and fast greedy algorithms, while retaining the accuracy rate in identifying the community membership of nodes. Taken together our results show that this new approach can be applied to speed up virtually any community detection method in dense complex networks, as it is the case of similarity networks.
\end{abstract}

%
%
%
%
%

\section{Introduction}

A myriad of real systems can be modeled as complex networks, where entities and their relationships are represented as nodes and edges, respectively. Examples of such systems are the Internet~\cite{net1,net2}, the WWW~\cite{www}, transport~\cite{road1,road2} and transmission systems~\cite{blackout}. Relevant in this modeling has been the ability of nodes to cluster into communities, defined as groups of strongly connected nodes with a few external links with the other nodes of the network. Various methods for detecting communities have been proposed~\cite{rico}, including waltkrap~\cite{wtrapref}, fast greedy~\cite{fref}, edge-betweenness~\cite{edgbtwref} and leading eigenvector~\cite{leadingref}. Unprecedented patterns of topological organization could be unveiled with communities being identified for metabolic, genetic, collaborative and social networks~\cite{generef,pathway,infor,jazz}.

Major issues for these methods are not only the accuracy but also the efficiency of the algorithm, since some real networks may comprise millions of nodes~\cite{webstructure,howpopular}. Actually, time efficiency is decisive for choosing the method for addressing a given problem as some methods become impractical for very large networks. This is the case of the edge-betweenness method, whose temporal complexity is $O(n^3)$ in the worst case. Perhaps because of the relevance of time efficiency, other important issues have been relatively neglected. An example is the applicability of standard methods in incomplete networks, i.e., networks with imprecise information, such as missing nodes or edges. To our knowledge, only a few studies have probed the efficiency of community detection methods in incomplete networks. In Ref.~\cite{1findingmissing} the authors focus on the predictability of missing edges, which is crucial for real networks resulting from incomplete experiments~\cite{margalita}. In information and social networks, for example, low-degree nodes are usually undiscoverable in crawling systems, while in protein interaction networks many edges may be unknown~\cite{reconstruction}.
Other related studies include the investigation of the robustness of communities when edge weights are varied and rewiring processes are applied~\cite{gfel,levina}.

In this paper, we evaluate the robustness of two methods in discovering communities in incomplete networks generated from sampling processes. As we shall show, these methods are robust even when several nodes are missing. Furthermore, we found out that the robustness seems to be weakly dependent on the method evaluated, but there is an important dependence on the network structure. More importantly, we show that robustness in detecting communities allows us to devise a strategy that improves the time performance, while keeping the accuracy of detecting communities in dense graphs such as similarity networks. One of the major advantages of the proposed strategy is that it can be applied to virtually all standard methods, since it relies on detecting communities in sampled networks.

\section{Methods} \label{metodos}

For the description of community detection methods, consider the following notation. A network is defined as $\mathcal{G}  = \{ \mathcal{V}, \mathcal{E} \}$, where $\mathcal{V}$ and $\mathcal{E}$ are respectively the set of nodes and edges. The connectivity is represented as an adjacency matrix $\mathcal{A} = \{a_{ij}\}$ with elements
\begin{equation}
a_{ij} = \left\{
\begin{array}{rl}
1, & \textrm{if $i$ and $j$ are linked},\\
0, & \textrm{otherwise.}\\
\end{array} \right.
\end{equation}
The degree of node $i$ is given by $k_i = \sum_j a_{ij}$. $\mathcal{D} = \{\delta_{ij}\}$ is the diagonal matrix. The element $\delta_{ij}$ is
\begin{equation}
\delta_{ij} = \left\{
\begin{array}{rl}
k_i, & \textrm{if } i = j,\\
0, & \textrm{otherwise.}\\
\end{array} \right.
\end{equation}
$\mathcal{P}_{ij} = \mathcal{D}^{-1} \mathcal{A} = \{p_{ij}\}$ is the Markovian adjacency matrix. Each element $p_{ij}$, defined as $p_{ij} = {a_{ij}}/{k_i}$, represents the probability of a random walker at node $i$ to reach node $j$ in the next time step.

The algorithms selected here to detect communities in sampled networks are the walktrap and fast greedy methods, which were chosen because they are suitable for weighted networks that are generated with our approach.

\subsection{Walktrap}

\begin{figure}
\centering
\includegraphics[width=0.85\textwidth]{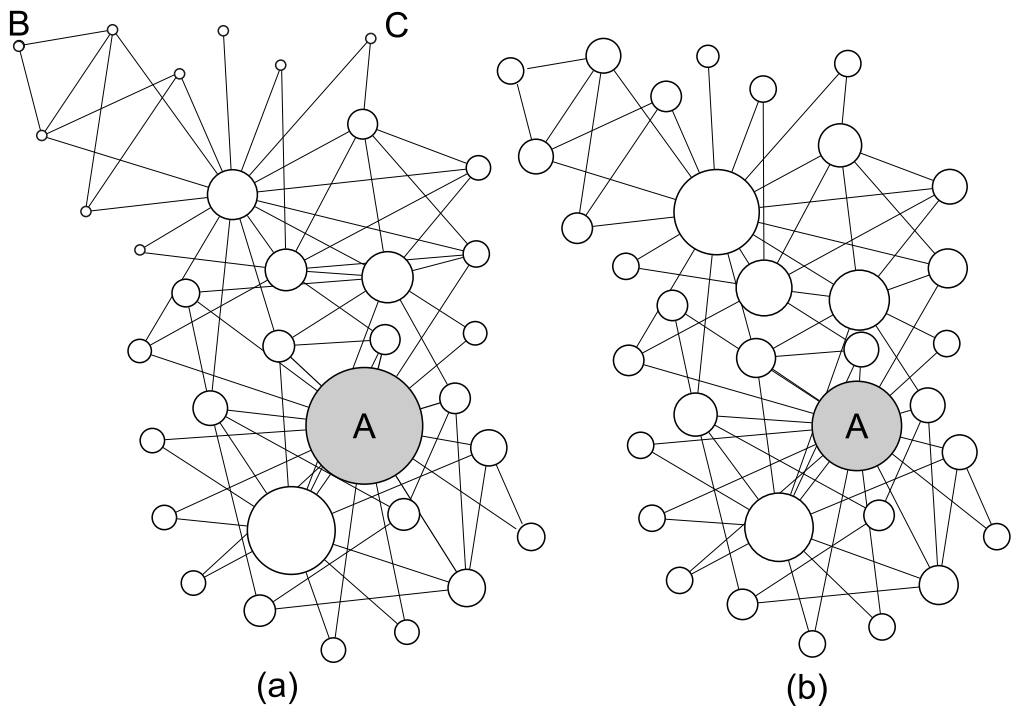}
\caption{\Red{Similarity between node ``A'' and the other nodes. The diameter of the nodes is proportional to their similarity with node ``A''. In (a) random walks of length $h=4$ were used and in (b) we used random walks of infinite length ($h \to \infty)$. In panel (a), nodes ``B'' and ``C'' are very dissimilar from ``A'' because their distance from ``A'' is large. In panel (b), similarities are assigned regardless of the distance from A. Actually, the only factor that matters in this case is the degree.}
\label{fig:rwalks}}
\end{figure}

The walktrap community detection method relies on random walks to split the network in natural partitions. At each time step, a particle moving on the network leaps to a neighboring node, which is chosen randomly. This process is repeated many times so that a Markov chain~\cite{markito} is generated. Here, the walker is allowed to leap onto a neighbor in fixed, discrete time steps. Random walks are used in walktrap to create a node similarity metric, which in turn is used to cluster nodes into communities. Two nodes $i$ and $j$ are considered similar if a random walk starting at $i$ accesses node $j$ many times. This similarity can be obtained analytically from the matrix $\mathcal{P}^t$, whose element $p_{ij}^{(t)}$ quantifies the probability of the walker to reach node $j$ (from node $i$) in $t$ steps. Each element $p_{ij}^{(t)}$ of $\mathcal{P}^{t}$ satisfies the relation $p_{ij}^{(t)} = k_j k_i^{-1} p_{ji}^{(t)}$. Therefore, if node $i$ is highly connected, it will reach node $j$ only a few times. Conversely, the higher the degree of node $j$ the higher is its probability to be reached from a random walk starting at any other node. In the steady state (i.e., in the limit as $t \to \infty$), the stationary probability $\pi_i \equiv \lim_{t \to \infty} p_{ij}^{(t)}$ $\forall i$ becomes:
\begin{equation}
     \pi_i \equiv \lim_{t \to \infty} p_{ij}^{(t)} = k_j / \sum_l k_l.
\end{equation}
Therefore, the parameter $t$ should not be much higher than the mixing time~\cite{markito} of $\mathcal{P}$, otherwise the likelihood $p_{ij}^{(t)}$ would reflect only the connectivity \Red{(see fig.~\ref{fig:rwalks}(b))}. In addition, $t$ should not take very low values because far distant nodes would be inaccessible \Red{(see fig.~\ref{fig:rwalks}(a))}.
%
%

Given the transition matrix $\mathcal{P}$, the distance $r_{ij}^{(t)}$ between nodes $i$ and $j$ is given by
\begin{equation} \label{sm1}
    r_{ij}^{(t)} = \sqrt{\sum_{l} \frac{ ( p_{il}^{(t)} - p_{jl}^{(t)} )^2 }{k_l} }
    = \| \mathcal{D}^{-\frac{1}{2}} \mathcal{P}_{i}^{(t)} - \mathcal{D}^{-\frac{1}{2}} \mathcal{P}_{j}^{(t)} \|,
\end{equation}
where $\mathcal{P}_{i}^{(t)}$ is the i-th row of $\mathcal{P}^t$. This metric can be generalized to measure the similarity $r_{\mathcal{C}_1\mathcal{C}_2}$ between two communities $\mathcal{C}_1$ and $\mathcal{C}_2$. Prior to the definition of $r_{\mathcal{C}_1\mathcal{C}_2}$, one needs to define the probability $\mathcal{P}_{\mathcal{C}_j}^{(t)}$ of a node $i \in \mathcal{C}$ to reach node $j \not\in \mathcal{C}$ in $t$ steps. This quantity is defined as
\begin{equation}
    \mathcal{P}_{\mathcal{C}_j}^{(t)} = \frac{1}{\|\mathcal{C}\|} \sum_{i \in \mathcal{C}} p_{ij}^{(t)},
\end{equation}
which represents the average likelihood of a node $i \in \mathcal{C}$ to reach a node $j \not\in \mathcal{C}$. With this definition, the distance between two communities is
\begin{eqnarray}
r_{\mathcal{C}_1\mathcal{C}_2} & = \sqrt{\sum_{l} \frac{ ( p_{\mathcal{C}_1l}^{(t)} - p_{\mathcal{C}_2l}^{(t)} )^2 }{k_l} } \nonumber \\
& = \| \mathcal{D}^{-\frac{1}{2}} \mathcal{P}_{\mathcal{C}_1}^{(t)} - \mathcal{D}^{-\frac{1}{2}} \mathcal{P}_{\mathcal{C}_2}^{(t)} \|.
\end{eqnarray}

\begin{figure}
\centering
\includegraphics[width=0.85\textwidth]{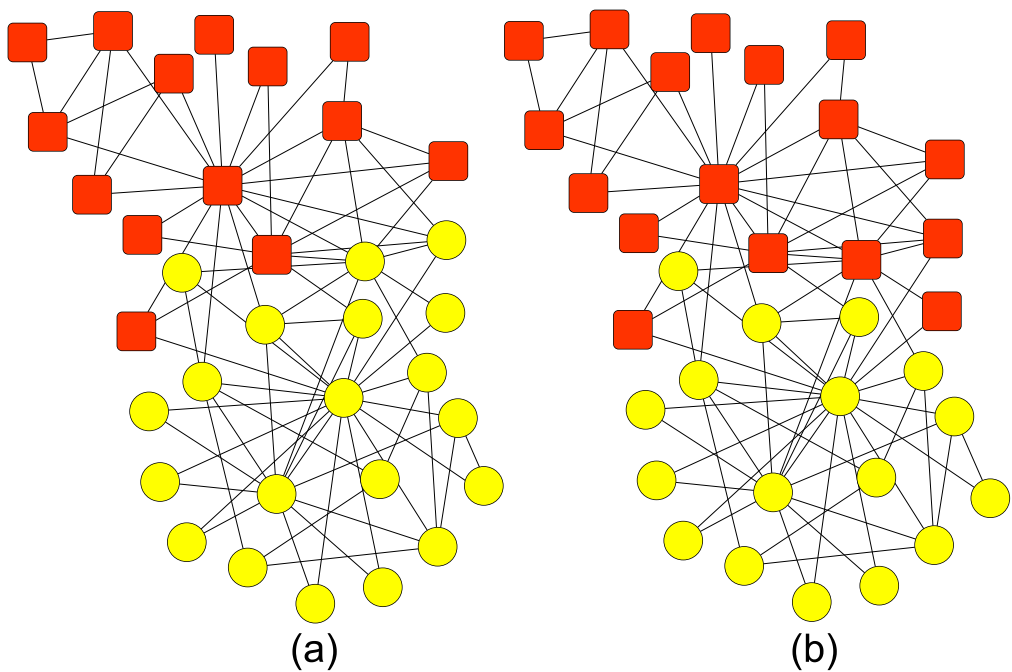}
\caption{\Red{Example of two communities detected with (a) walktrap; and (b) fast greedy methods. The network employed in this example is the social network of friendships between 34 members of a karate club at a US university in the 1970s~\cite{karatecit}.}
\label{fig:karate}}
\end{figure}

After computing all pairs of distances between communities, the walktrap method follows an agglomerative approach based on the Wards method~\cite{ward}. Initially, each node represents a community. Two communities $\mathcal{C}_1$ and $\mathcal{C}_2$ are merged if the new partition minimizes $\sigma$, the squared distances between nodes and their respective communities:
\begin{equation}
    \sigma_l =  \sum_{\mathcal{C}} \sum_{i \in \mathcal{C}} r^2_{i\mathcal{C}},
\end{equation}
where $r_{i\mathcal{C}} \equiv r_{\{i\}\mathcal{C}}$. Then, a new community $\mathcal{C}_{n+1} = \mathcal{C}_1 \cup \mathcal{C}_2$ arises and the old partition $P_l$ becomes $( P_l~\setminus~\{\mathcal{C}_1,\mathcal{C}_2\}) \cup \{\mathcal{C}_{n+1}\}$ ). Finally, the process is repeated until the expected number of communities is obtained. The detection of two communities using this method is \Red{illustrated in fig.~\ref{fig:karate}(a).}

Note that the distance $r_{ij}^{(t)}$, as defined in eq. (\ref{sm1}), has a strong relationship with the spectra of $\mathcal{P}$. More specifically, $r_{ij}^{(t)}$ can be rewritten as
\begin{equation} \label{eigen}
    r_{ij}^{(t)} =  \Bigg{[} \sum_{\alpha=2} \lambda_\alpha^{2t} ( v_\alpha(i) - v_\alpha(j) )^2 ) ^ 2 \Bigg{]}^{\frac{1}{2}},
\end{equation}
where $\lambda_\alpha$ and $v_\alpha$ are respectively the eigenvalues and eigenvectors of $\mathcal{P}$. In view of this formulation in terms of graph spectra, $r^{(t)}_{ij}$ in eq. (\ref{eigen}) may be defined to consider different weighting for distinct eigenvalues, allowing thus the use of continuous random walks~\cite{weiss}. This generalization is achieved with the following relation %
\begin{eqnarray}
    r_{ij}^2 & = \sum_{\alpha = 2}^N \Bigg{(} \sum_{l=0}^\infty c_l \lambda_\alpha^l \Bigg{)}^2  ( v_\alpha(i) - v_\alpha(j) )^2 \nonumber \\
    & = \| \mathcal{D}^{-\frac{1}{2}} \tilde{\mathcal{P}}_{i}^{(t)} - \mathcal{D}^{-\frac{1}{2}} \tilde{\mathcal{P}}_{j}^{(t)} \|,
\end{eqnarray}
where
\begin{equation}
\tilde{\mathcal{P}}_{i} = \sum_{l=0}^{\infty} c_l \mathcal{P}_i^{(l)}.
\end{equation}

\subsection{Fast greedy}

Similarly to the walktrap community detection method, the fast greedy algorithm is also based on hierarchical agglomerative clustering. Initially, each node represents a community. As the algorithm is progressively applied, similar nodes are joined into communities (the similarity is established according to a given criterion) until all nodes belong to a same giant community, thus completing the dendrogram. To join two nodes, the algorithm uses the modularity $Q$, which measures the number of intra-community edges that are higher than the expected by chance. The quantity $Q$ is
\begin{eqnarray} \label{eeq}
    Q & = \sum_i (e_{ii} - a_i^2) \nonumber \\
    & = \frac{1}{2m} \sum_i \sum_j \Bigg{(} a_{ij} - \frac{k_i k_j}{2m} \Bigg{)} \epsilon(g_i,g_j),
\end{eqnarray}
where $e_{ij}$ is the fraction of edges linking nodes in community $i$ to those in community $j$, $a_i = \sum_j e_{ij}$, $m = 1/2 \sum k_i$, $g_i$ is the community to which node $i$ belongs and
\begin{equation*}
\epsilon(g_i,g_j) = \left\{
\begin{array}{rl}
1, & \textrm{if $g_i = g_j$},\\
0, & \textrm{otherwise.}\\
\end{array} \right.
\end{equation*}
More specifically, the fast greedy algorithm joins two communities provided that $\Delta Q = e_{ij} + e_{ji} - 2a_ia_j = 2(e_{ij} -a_ia_j)$ is  maximized. \Red{Note that eq. (\ref{eeq}) can be straightforwardly computed in a weighted network provided that it has been mapped to a multigraph~\cite{nweighted}}.

It is worth noting that it is not necessary to check all possible joining possibilities since only the junction of neighboring communities (i.e. communities with at least one edge linking two of their nodes) is able to increase $Q$. An example of a network with two communities identified with the fast greedy \Red{method is depicted in fig~\ref{fig:karate}(b)}.

\section{Results and discussion} \label{effect}

\subsection{Community detection in incomplete networks} \label{effectsec}

The ability of the community detection algorithms to find natural clusters in incomplete networks was tested with the following methodology. We started with toy networks $\mathcal{G}=\{\mathcal{V},\mathcal{E}\}$, henceforth referred to as original networks, generated according to the procedures described in Refs.~\cite{bench1,bench2}. The following parameters were employed: $N$, the number of nodes; $\langle k \rangle = 1/N \sum k$, the average degree and $\mu$, the mixing parameter (quantifies the fraction of links that are placed outside the community of the node). \Red{According to previous studies~\cite{1findingmissing,2fmiss}, we chose $\mu = 0.30$. Values of mixing parameter above $\mu \simeq 0.3$ destroy the modular structure of the network. This is apparent when one observes, for example,  that the normalized mutual information ($\Gamma$)~\cite{refjstat} obtained with the walktrap in $\mathcal{N}_b$ for $\mu=0.35$ and $\mu=0.40$ were $\Gamma=0.589$ and $\Gamma=0.001$, respectively.}

To create an incomplete version $\mathcal{G}' = \{\mathcal{V}' \in \mathcal{V},\mathcal{E}'\}$ of $\mathcal{G}$, the nodes in $\mathcal{G}$ were randomly sampled with sampling rate $\mathcal{S}$. The unweighted connectivity matrix $\mathcal{A}$ becomes an weighted matrix $A'$ such that $a_{ij} \mapsto d_{ij}^{-1}, \forall (i,j) \in V'$, where $d_{ij}$ represents the length of the shortest path linking nodes $i$ and $j$ in $\mathcal{G}$. \Red{This approach relying oh the random selection of nodes is similar to the one employed in the study performed in~\cite{newref92}, which investigated the convergence of spectral clustering methods (in the machine learning context) for increasing sample size. As we shall show, the main focus of our study is to use the random selection of nodes for devising a more efficient version of traditional community detection methods.}
In the experiments, we used the following networks: 

\begin{itemize}
  \item Network $\mathcal{N}_a$: $N=512$ and $\langle k \rangle = 32$,
  \item Network $\mathcal{N}_b$: $N=1,024$ and $\langle k \rangle = 48$,
  \item Network $\mathcal{N}_c$: $N=2,048$ and $\langle k \rangle = 96$,
  \item Network $\mathcal{N}_d$: $N=512$ and $\langle k \rangle = 64$,
  \item Network $\mathcal{N}_e$: $N=1,024$ and $\langle k \rangle = 96$,
  \item Network $\mathcal{N}_f$: $N=2,048$ and $\langle k \rangle = 128$.
\end{itemize}


To compute the accuracy rate in identifying the community structure of complex networks, consider the following definition. Let $c_i^{(r)}$ and $c_i^{(m)}$ be the community associated with node $i$ in the reference network and the community associated with the same node by one of the community detection algorithms. Note that, to quantify the accuracy rate it is not enough to compare $c_i^{(r)}$ and $c_i^{(m)}$, because the labeling scheme employed by the algorithm might be different from the one employed to describe the communities in the reference network.
For example, if $c^{(r)} = \{1,1,2,2\}$ and $c^{(m)} = \{2,2,1,1\}$, the accuracy rate obtained from a straightforward comparison would be $\Gamma=0$, even though the communities are equivalent. To consider all possible labeling schemes for $c^{(m)}$, it is possible to apply the operator $\mathcal{L}$, which maps $c^{(m)}$ to every possible labeling scheme. According to the previous example, the application of $\mathcal{L}$ to $c^{(m)}$ would lead to the mapping $\mathcal{L}(c^{(m)}) = \{ \{2,2,1,1\}, \{1,1,2,2\} \}$. Equivalently, $c^{(m,1)} =  \{2,2,1,1\}$ and $c^{(m,2)} = \{1,1,2,2\}$. Thus, the accuracy rate can be defined as
\begin{equation} \label{aceq}
	\alpha = \max_j \sum_j  \sum_i \delta(c_i^{(r)},c_i^{(m,j)}),
\end{equation}
where $\delta(x,y) = 1$ if $x=y$ and $\delta(x,y) = 0$ if $x \neq y$.
\Red{Even though the measurement defined in equation \ref{aceq} is able to capture the quality of the found partition, it depends upon the computation of several permutations. To avoid such costly computation, we used instead the normalized mutual information, which can be computed as
\begin{equation}
    \Gamma(A,B) = \frac{-2 \sum_{i=1}^{c_A} \sum_{j=1}^{c_B} n_{ij} \log( n_{ij} n / n_{i:}n_{j:} )}{\sum_{i=1}^{c_A} n_{i:} \log(n_{i:}/n) + \sum_{j=1}^{c_B} n_{:j} \log(n_{i:}/n)  },
\end{equation}
where $c_A$ and $c_B$ are respectively the number of real and found communities, $n_{ij}$ is the number of nodes in the original community $i$ that appear in the found community $j$, $n_{i:} = \sum_j n_{ij}$ and $n_{:j} = \sum_i n_{ij}$. The normalized mutual information has been shown to perform well in practice and for this reason has been used in the community detection field~\cite{refjstat,refjain}.}

\begin{figure}
\centering
\includegraphics[width=0.75\textwidth]{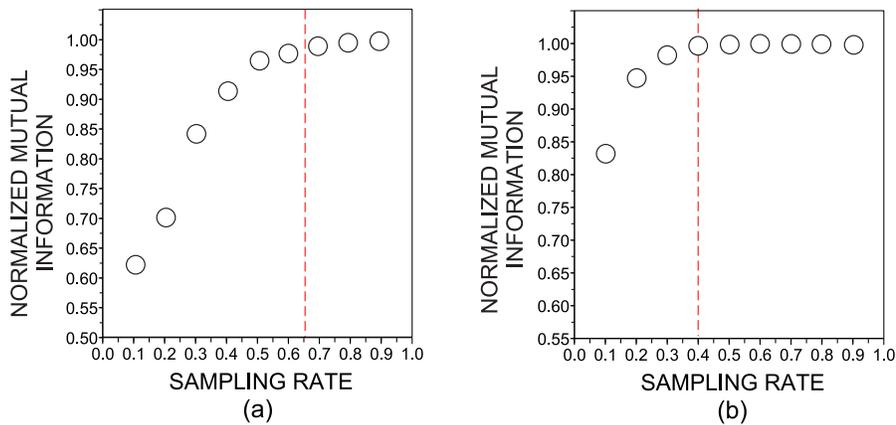}
\caption{\Red{Dependence of the accuracy with the walktrap algorithm on the sampling rate in network (a) $\mathcal{N}_a$; and (b) $\mathcal{N}_d$. The vertical dashed lines represent the threshold $\varsigma$. As an increasing number of nodes are removed from the network, the ability to detect communities decreases. Note that all networks are robust to node removal (in the sense that they keep their community structure) provided that the sampling rate is above a given threshold. Similar results were obtained for the other synthetic networks.}
}
\label{fig:walktrap_sampling}
\end{figure}

The ability to detect community structures in fig.~\ref{fig:walktrap_sampling} using the walktrap algorithm is similar for all the networks considered. The performance is very high for incomplete networks with spurious edges (i.e., when the sampling is performed with sampling rate $\mathcal{S} = 1$), as revealed by value of normalized mutual information $\Gamma$ above $0.95$. When $\mathcal{S}$ decreases and therefore less nodes are left in the incomplete network, the ability to detect communities diminishes, as one should expect. Interestingly, for all networks, there exists a threshold $\mathcal{S} = \varsigma$ discriminating two regimes. When $\mathcal{S} \leq \varsigma$, the organization in communities disappears rapidly as $\mathcal{S}$ decreases. In contrast, for $\mathcal{S} > \varsigma$, the community structure seems to be maintained in spite of the removal of many nodes. This is apparent for network $\mathcal{N}_d$ (see fig. \ref{fig:walktrap_sampling}(b)), for example. Even with 60\% of the nodes being discarded ($\mathcal{S} = 0.40$), the communities are well distinguished from each other. Also, the degree of connectivity $\langle k \rangle$ affects $\varsigma$. The increase in $\langle k \rangle$ causes the network to be more robust so that the modular organization does not disappear at all, an effect that becomes even more evident by comparing $\mathcal{N}_a$ ($\langle k \rangle = 32$) and $\mathcal{N}_d$ ($\langle k \rangle = 64$). While in the former the threshold is $\varsigma \sim 0.65$, the latter displays a threshold $\varsigma \sim 0.40$.

\begin{figure}
\centering
\includegraphics[width=0.75\textwidth]{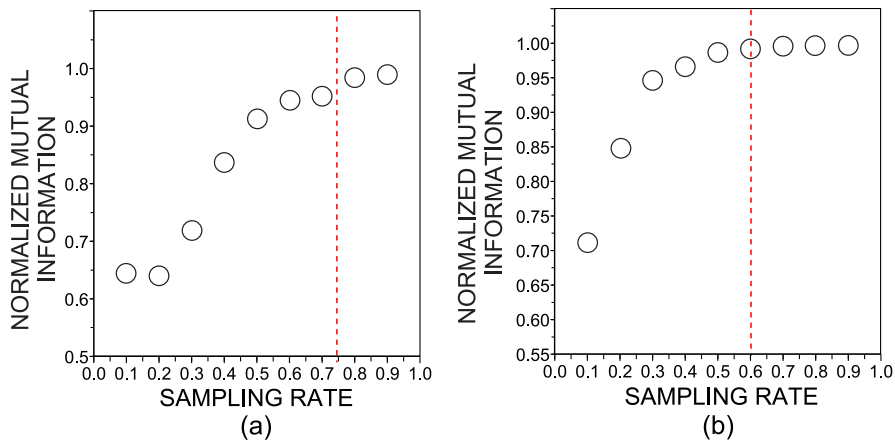}
\caption{\Red{Dependence on the sampling rate for the accuracy using the fast greedy algorithm for network (a) $\mathcal{N}_a$; and (b) $\mathcal{N}_d$. The vertical dashed lines represent the threshold $\varsigma$. The ability to discriminate communities drops as an increasing number of nodes are discarded. Note that all networks are robust to node removal (in the sense that they keep their community structure) provided that the sampling rate is above a given threshold. Similar results were obtained for the other synthetic networks.}
\label{fig:greedy_sampling}}
\end{figure}

Fig.~\ref{fig:greedy_sampling} displays how the \Red{normalized mutual information} varies with the sampling rate for the fast greedy method used to detect communities. The results are essentially similar to those of the walktrap method in fig.~\ref{fig:walktrap_sampling}. The fast greedy method performs well when $\mathcal{S} = 1$ (i.e., when no node is removed), just as in the walktrap. As nodes are removed with a sampling rate $\mathcal{S} < \varsigma$, the \Red{normalized mutual information} $\Gamma$ decreases at a low rate. The values of $\varsigma$ for both methods are similar, suggesting a stronger dependence on network topology. The robustness of the network (in the sense that the community structure is maintained) increases with the average connectivity $\langle k \rangle$, as indicated by comparing $\mathcal{N}_a$ and $\mathcal{N}_d$ in fig.~\ref{fig:greedy_sampling}.

All in all the results reveal that the community detection methods evaluated are robust for they are able to identify the modular organization even when many nodes from the original network are removed. Hence if we are interested in finding the community to which only a few nodes belong, we can choose to deliberately eliminate the other nodes from the analysis. Provided that the sampling rate is sufficiently large (i.e., $\mathcal{S} > \varsigma$), high accuracy can be achieved with a gain in performance, since computation in smaller networks implies a decrease in computational cost. This idea of detecting community in sampled networks with a gain in temporal performance serves as motivation to the proposed method described below.

\subsection{Fast community detection via sampling processes in synthetic networks} \label{fast}

\begin{figure}
\centering
\includegraphics[width=1\textwidth]{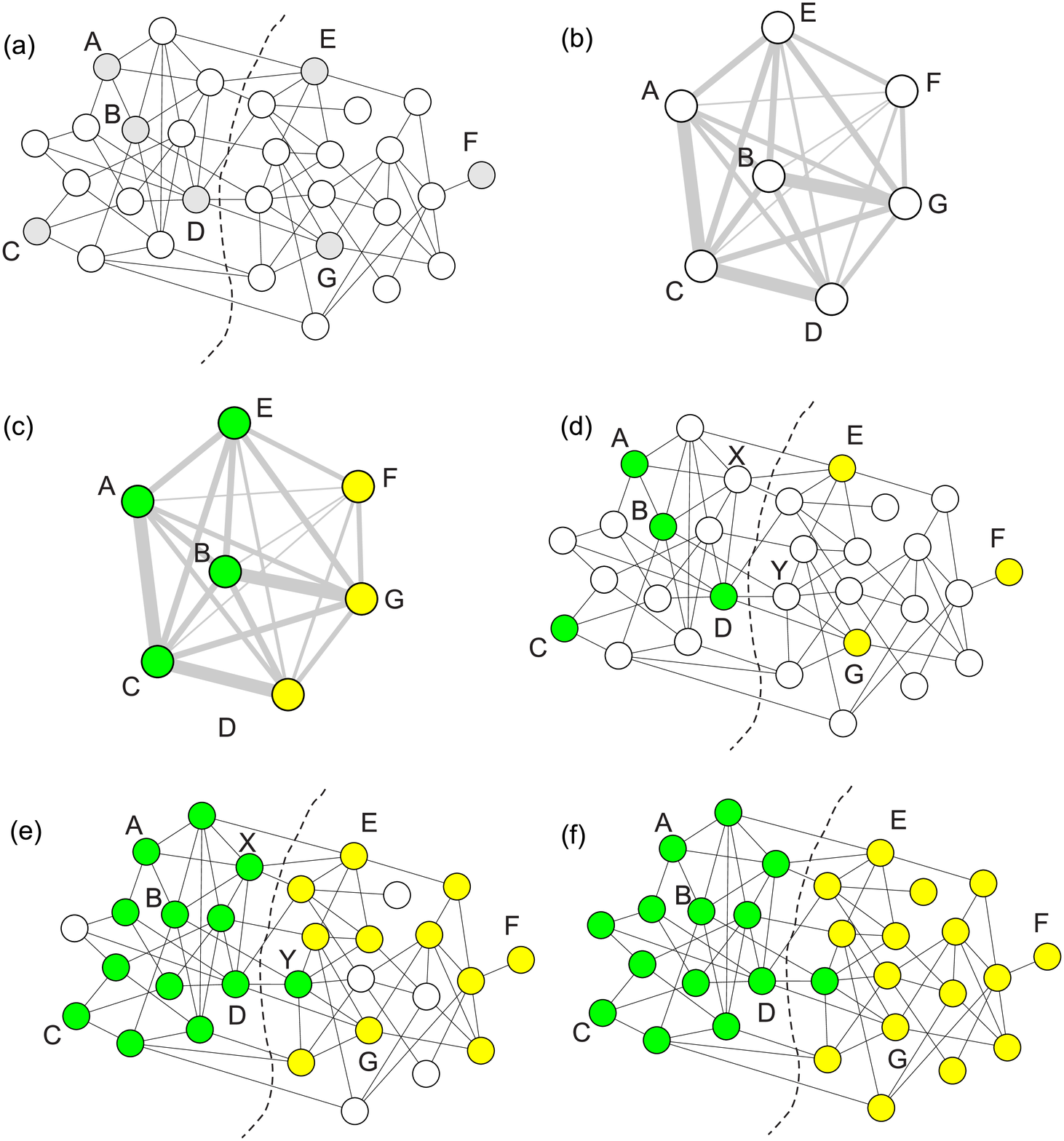}
\caption{\Red{Evolution of the community detection method based on the analysis of incomplete networks. (a) (step 1) Sampling in the original network. (b) (step 2) Construction of the incomplete network (the edge thickness is proportional to the strength of the links). (c) (step 3) Community detection in the incomplete network. (d) (step 4) Transference of the memberships obtained in the incomplete network to the original network. (e) (step 5) Label propagation in the original network. (f) (step 6) Repetition of step 5 until all nodes are classified}.
\label{fig:awesome_image} }
\end{figure}

\Red{The finding that the community structure is maintained in incomplete networks derived from a random sampling process with a sampling rate $\mathcal{S} > \varsigma$ motivated us to devise a method to decrease the computational cost of the walktrap and fast greedy methods. As we shall show, this gain in time performance has a low impact on the quality of the found partition provided that the network is sufficiently connected.}
The proposed algorithm initially randomly chooses a set $\mathcal{V}' \in \mathcal{V}$ such that $\| \mathcal{V'} \|~\| \mathcal{V} \|^{-1} = \mathcal{S}$. Then the selected nodes are connected with weights $a_{ij}' = d_{ij}^{-1}$, where $d_{ij}$ is the length of the shortest path linking nodes $i$ and $j$ in the complete (not sampled) network. Note that this procedure coincides with the one adopted to form incomplete networks in the previous section. In the next step, communities are discovered using any standard method. Then the membership assigned for each node in the sampled network is mapped to the corresponding node in the original network. To assign the membership of the remaining nodes in $\mathcal{V}$, a voting strategy over the neighbors is adopted. If most of the neighbors belong to the community $\mathcal{C}$, then $\mathcal{C}$ is assigned to that node. In case of ties, the decision is postponed to the next iteration. This process is repeated until all nodes have been classified. The overall process can be summarized in 6 steps:

\begin{enumerate}

  \item {\bf Step 1}: Select randomly a set of nodes from the original network.

  \item {\bf Step 2}: Create an incomplete network whose edges weights are inversely proportional to the distances in the original network. \ \ \

  \item {\bf Step 3}: Identify the communities in the simplified network using any standard community detection method (e.g., walktrap or fast greedy).

  \item {\bf Step 4}: Transfer the memberships obtained in the incomplete network to the original network.

  \item {\bf Step 5}: Propagate labels according to a voting strategy over neighbors.

  \item {\bf Step 6}: Repeat step 5 until all nodes have been classified.

\end{enumerate}
\Red{The process of detecting communities with the above method is illustrated in the original toy network displayed in fig.~\ref{fig:awesome_image}(a). The two communities are divided by a dashed line. Highlighted nodes represent those selected randomly. Initially, an incomplete network comprising the nodes  randomly selected from the original network is formed (fig.~\ref{fig:awesome_image}(b)). After detecting the communities in the incomplete network (fig.~\ref{fig:awesome_image}(c)), the membership of each node is transferred to the original network, giving rise to the configuration depicted in fig.~\ref{fig:awesome_image}(d). Then the label propagation phase takes over until all nodes are classified. The result of the first iteration is displayed in fig.~\ref{fig:awesome_image}(e). Note that node $X$ has been classified as belonging to the `green' community because it is connected to two nodes belonging to the 'green' community and just one belonging to the 'yellow' community. On the other hand, node $Y$ was incorrectly classified as 'green' because it is connected to another 'green' node. The final configuration after the second iteration is shown in fig.~\ref{fig:awesome_image}(f).}



\begin{figure}
\centering
\includegraphics[width=1\textwidth]{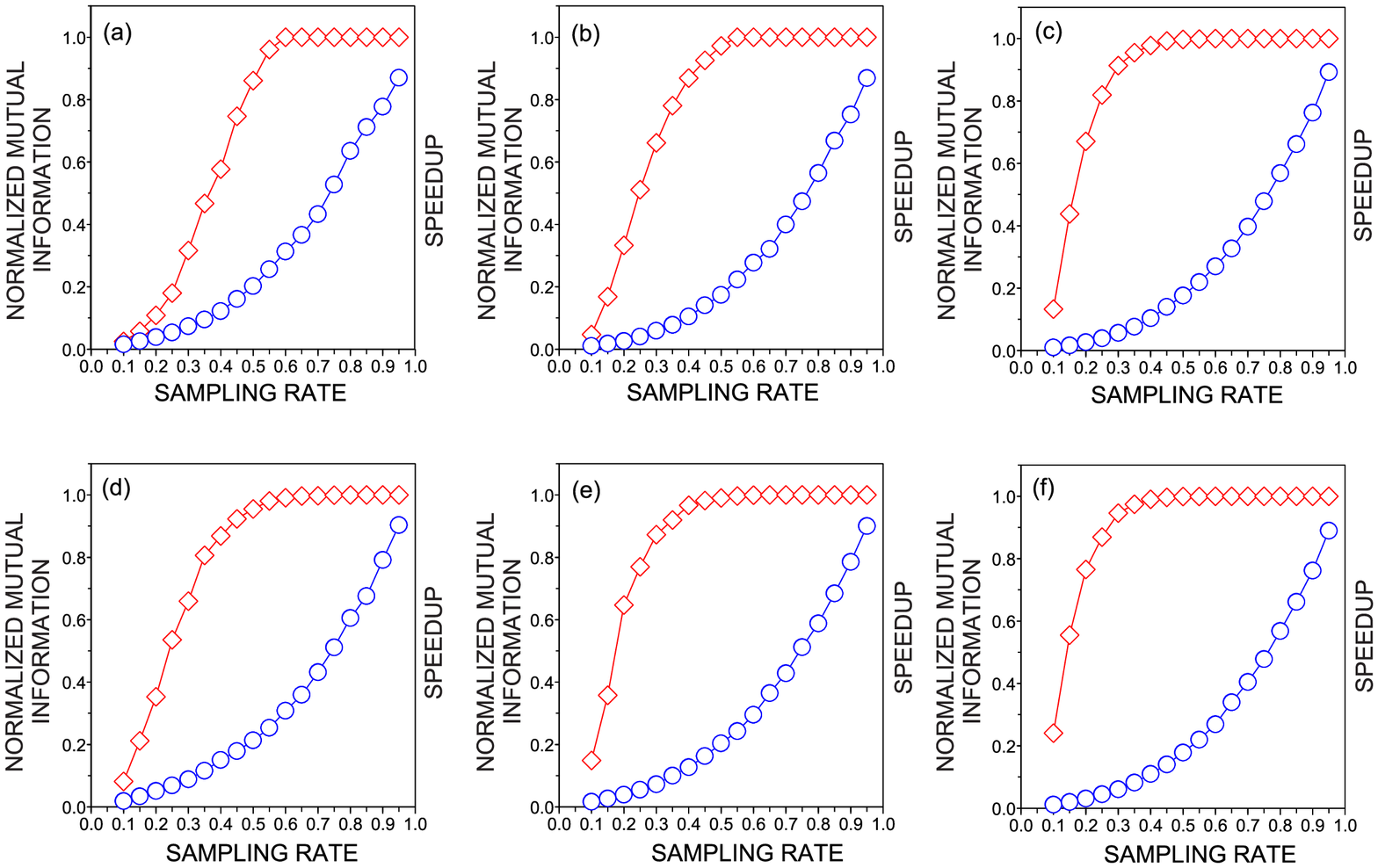}
\caption{\Red{Dependence of the normalized mutual information and speedup with the sampling rate $\mathcal{S}$ in (a) $\mathcal{N}_a$; (b) $\mathcal{N}_b$; (c) $\mathcal{N}_c$; (d) $\mathcal{N}_d$, (e) $\mathcal{N}_e$ and (f) $\mathcal{N}_f$. The communities were identified with the \emph{walktrap} algorithm applied in step 3.
The standard deviation obtained for the normalized mutual information across distinct sampling networks is shown in fig.~S1 of the Supporting Information, which is available at \url{https://dl.dropboxusercontent.com/u/2740286/ssi.pdf}.}}
\label{fig:awesome_image_wk}
\end{figure}

\begin{figure}
\centering
\includegraphics[width=1\textwidth]{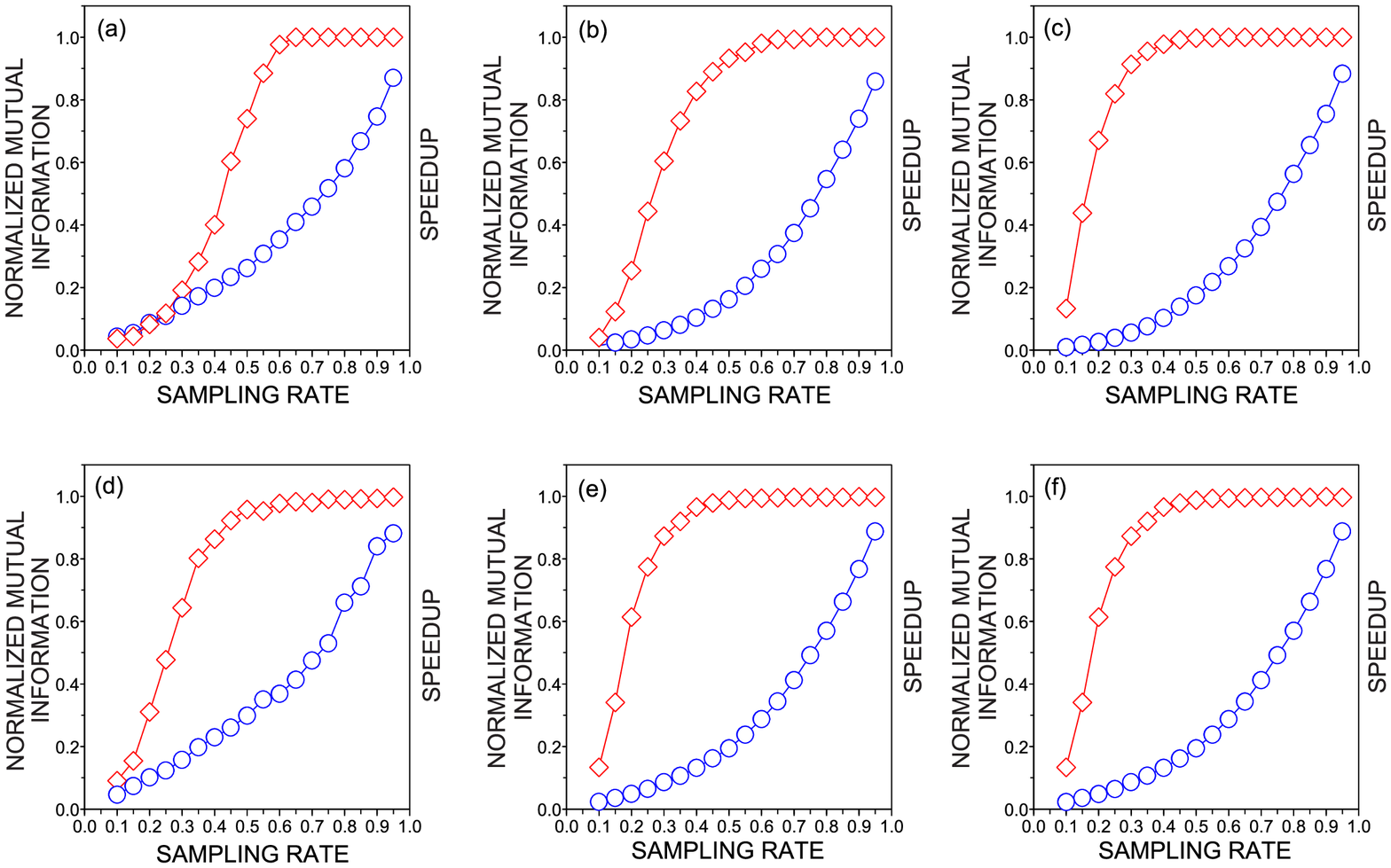}
\caption{\Red{Dependence of the normalized mutual information and speedup with the sampling rate $\mathcal{S}$ in (a) $\mathcal{N}_a$; (b) $\mathcal{N}_b$; (c) $\mathcal{N}_c$; (d) $\mathcal{N}_d$, (e) $\mathcal{N}_e$ and (f) $\mathcal{N}_f$. The communities were identified in step 3 with the \emph{fast greedy} algorithm.
The standard deviation obtained for the normalized mutual information across distinct sampled networks is shown in fig.~S2.
}}
\label{fig:fasty}
\end{figure}

\Red{The efficiency of the proposed technique was verified in the networks $\mathcal{N}_a$ - $\mathcal{N}_f$. The results obtained using the walktrap method in step 3 is displayed in fig.~\ref{fig:awesome_image_wk}. In each subplot, the upper curve refers to the \Red{normalized mutual information} $\Gamma$ in assigning communities, while the bottom one shows the normalized processing time -- the speedup (i.e. the time spent in performing the six steps divided by the time spent by the community detection method running directly on the original network). Interestingly, the accuracy rates after step 6 are similar to those in figs. \ref{fig:walktrap_sampling} and \ref{fig:greedy_sampling}, thus indicating that the accuracy of our method strongly depends on the ability to detect the communities in the sampled, incomplete networks (step 3). Provided that this detection is correct, the membership labels are propagated with minimum error. The curves of time performance reveal that it is feasible to achieve a high accuracy rate while improving time performance. For instance, in network $\mathcal{N}_b$ our method reaches a value of $\Gamma > 0.90$ and increases time performance in about 60\%. The comparison between $\mathcal{N}_b$ and $\mathcal{N}_e$ shows that the proposed method is even more effective when the average connectivity $\langle k \rangle$ of the original network takes high values. While a sampling rate of 30\% yields an partition with $\Gamma \simeq 0.68$ in $\mathcal{N}_b$, the same sampling rate yields an accuracy rate of $\Gamma \simeq 0.88$ in network $\mathcal{N}_e$. In the latter, our method runs around 10 times faster than the same algorithm running on the original network. With regard to the fastgreedy method, similar results were obtained (see fig.~\ref{fig:fasty})}.


\Red{To probe whether the community structure found after removing some nodes of the original network is significant, the following experiment was carried out. For each sampled network created in step 2, we generated 20 equivalent randomized versions. We then identified the communities in these random networks in order to verify if the generated partition is as accurate as the one generated with the sampled networks obtained in step 2. The results comparing sampled and randomized versions of $\mathcal{N}_b$ are shown in Table \ref{fastrandom}. Note that the normalized mutual information obtained in sampled networks is much larger than the one obtained in random networks, thus confirming the significance of the found communities. Similar results have been found for the other synthetic networks (results not shown).}

\begin{table}[h]
\caption[]{\label{fastrandom}\Red{Normalized mutual information obtained for sampled networks ($\Gamma^{(\textrm{s})}$) and their randomized versions ($\Gamma^{(\textrm{r})}$). The average and the standard deviation obtained in the randomized versions are represented respectively by $\langle \Gamma^{(\textrm{r})} \rangle$ and $\Delta \Gamma^{(\textrm{r})}$. The results were obtained with the fast greedy method in the network $\mathcal{N}_b$.}}
\centering
\begin{tabular}{@{}|c|c|c|c|}
\hline
$\mathcal{S}$ & $\Gamma^{(\textrm{s})}$ & $\langle \Gamma^{(\textrm{r})} \rangle$ & $\Delta \Gamma^{(\textrm{r})}$ \\ \hline
0.10  &  0.106 &  0.012 &  0.023 \\
0.15  &  0.645 &  0.007 &  0.008 \\
0.20  &  0.507 &  0.005 &  0.011 \\
0.25  &  0.730 &  0.003 &  0.003 \\
0.30  &  0.720 &  0.005 &  0.008 \\
0.35  &  0.869 &  0.002 &  0.002 \\
0.40  &  0.960 &  0.002 &  0.003 \\
0.45  &  0.944 &  0.003 &  0.003 \\
0.50  &  0.982 &  0.003 &  0.003 \\
0.55  &  1.000 &  0.001 &  0.002 \\
0.60  &  1.000 &  0.001 &  0.001 \\
0.65  &  1.000 &  0.001 &  0.001 \\
0.70  &  0.972 &  0.002 &  0.001 \\
0.75  &  1.000 &  0.001 &  0.001 \\
0.80  &  1.000 &  0.001 &  0.001 \\
0.85  &  1.000 &  0.001 &  0.002 \\
0.90  &  1.000 &  0.001 &  0.001 \\
0.95  &  1.000 &  0.001 &  0.001 \\
\hline
\end{tabular}
\end{table}

\begin{figure}
\centering
\includegraphics[width=0.75\textwidth]{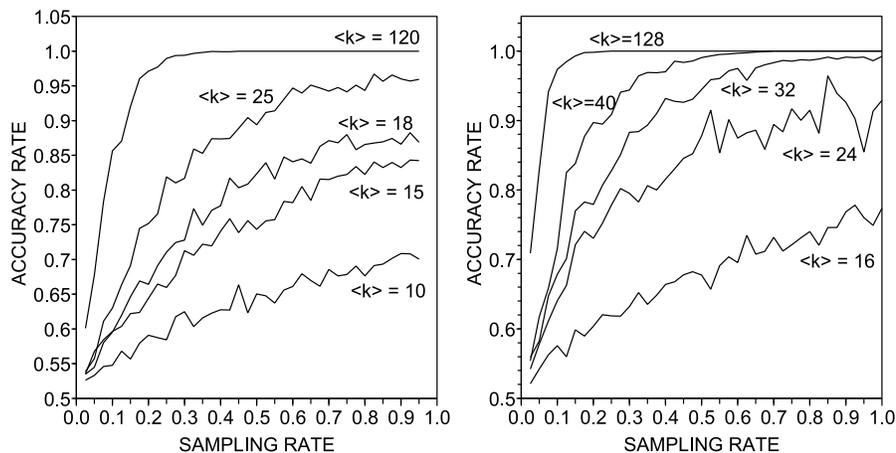}
\caption{{Dependence of $\Gamma$ obtained with the walktrap method as the sampling rate varies for networks comprising (a) $N= 512$ nodes; and (b) $N=1,024$ nodes.}}
\label{fig:graus_wt}
\end{figure}

In the light of the behavior displayed in figs.~\ref{fig:awesome_image_wk} and \ref{fig:fasty}, the average connectivity $\langle k \rangle$  seems to play a crucial role on the curves for accuracy versus sampling rates. A more detailed analysis of the relationship between the sampling rate $\mathcal{S}$ and accuracy rate $\Gamma$ was conducted on networks $\mathcal{N}_a$ and $\mathcal{N}_b$, with the results for the walktrap being shown in fig.~\ref{fig:graus_wt}.
It is clear that the tuning of $\langle k \rangle$ affects the threshold $\mathcal{S}$. Whenever $\langle k \rangle$ takes sufficiently low values (e.g. $\langle k \rangle = 10$ in fig.~\ref{fig:graus_wt}(a)), the community structure fades away even with high sampling rates. These results suggest that the strategy developed here is especially useful when the original network is very connected. Actually, our method is most suitable to detect communities in weighted, complete networks~\cite{xico}, for the sampling process ensures that both the number of nodes and edges decreases, thus assuring an enhancement in time performance.

The robustness of the network was also studied in networks comprising four communities. In this case, we noted that the quality of the partitions decreased after the removal of a few nodes (see fig.~S3). Therefore, in networks with a high number of overlapping communities, more efficient sampling node should be considered.

\subsection{Fast community detection via sampling processes in real networks} \label{realtest}

\Red{To complement the investigation of the properties of the proposed methodology to detect modular structures, we verified the influence of sampling nodes on the discriminability of communities in real networks. The following similarity networks were studied: the email network ($N=1,133$ and $m=5,452$, see~\cite{emailnetwork}) and the network of political blogs ($N=1,490$ and $m=16,715$, see~\cite{lada}). The results are shown in fig. \ref{real_net}. In this figure, the relative accuracy represents the fraction $\Gamma^{(S)}/\Gamma^{(O)}$, where $\Gamma^{(S)}$ and $\Gamma^{(O)}$ are the normalized mutual information obtained in the sampled and original networks, respectively. The email network turned out to be less resilient than the other synthetic networks studied in Section \ref{fast}. This is apparent when one notes that the accuracy decreases even when a small amount of nodes are removed. In this case, the walktrap method seems to be more robust than the fast greedy. The network of political blogs, on the other hand, displayed a more robust behavior for the accuracy as more nodes are removed. Interestingly, the relative accuracy remains high when about 50\% of the nodes are disregarded. Concerning the variability of the normalized mutual information across distinct sampled networks, we observed low values of coefficient of variation (see fig.~\ref{real_net}). In general, as expected, the highest values of variability occurred for the lowest values of the sampling rate $\mathcal{S}$.}


\begin{figure}
\centering
\includegraphics[width=0.65\textwidth]{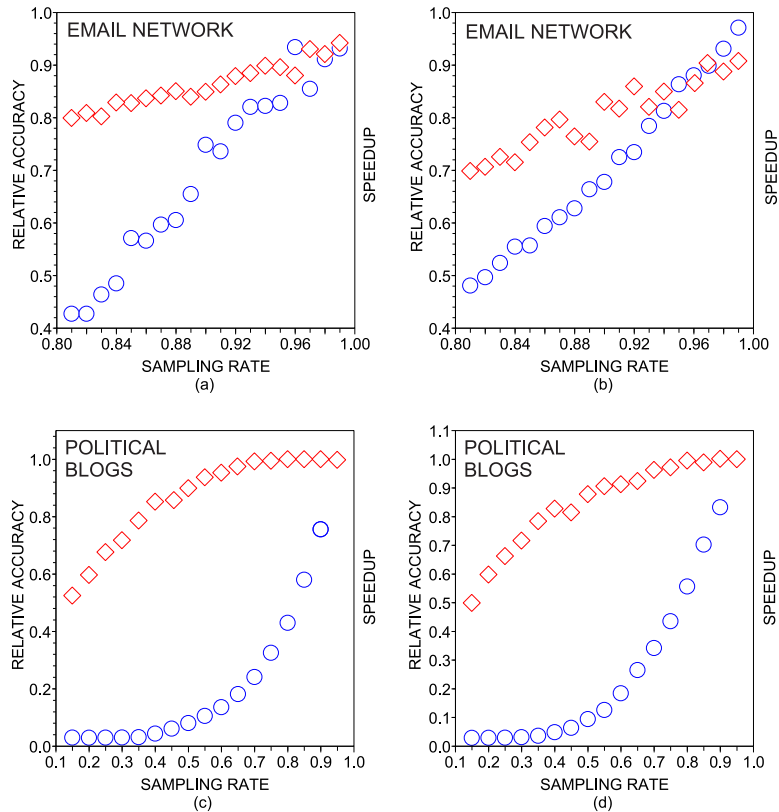}
\caption{\Red{Relative accuracy and speedup obtained in the network of messages and political blogs. The walktrap was employed in (a) and (c); and the fastgreedy was employed in (b) and (f). The network of political blogs seems to be more resilient than the email network. In the former, high values of relative accuracy were observed even when a large number of nodes were removed.}
}
\label{real_net}
\end{figure}

\begin{figure}
\centering
\includegraphics[width=0.65\textwidth]{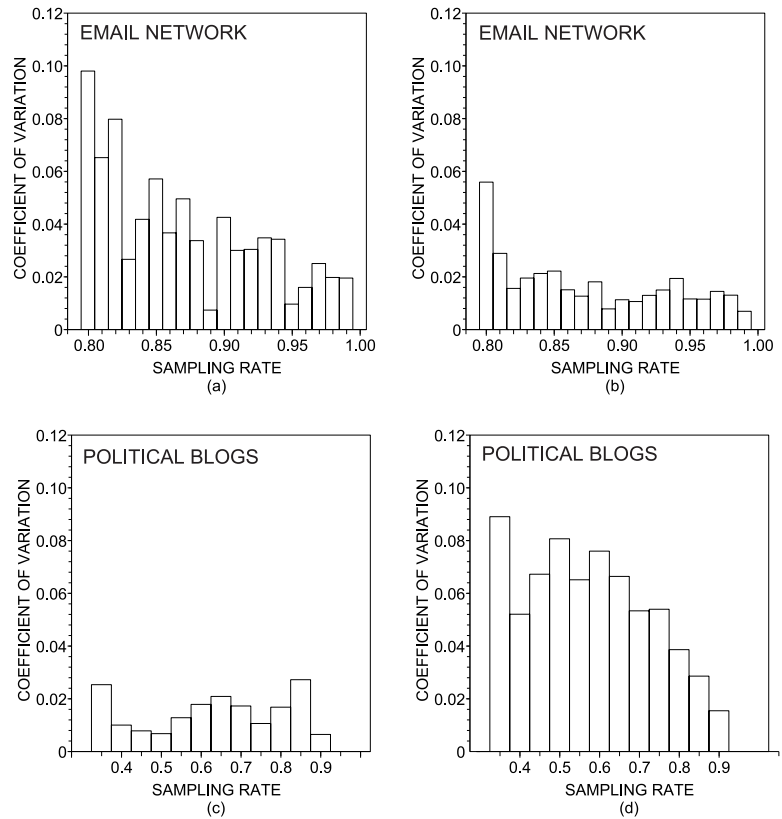}
\caption{\Red{Coefficient of variation of the normalized mutual information obtained in the network of messages and political blogs. The walktrap was employed in (a) and (c); and the fastgreedy was employed in (b) and (f). Low values of variability were observed in all cases.}}
\label{real_net}
\end{figure}
\Red{In this section, as proof of principle, we verified that our method can be applied in two real networks. However, further studies should clarify which conditions should be fulfilled so that the gain in performance via sampling nodes still provides good partitions. As noted for the synthetic networks, the gain in performance may depend upon the mixing parameter, the average connectivity and other factors. For this reason, we believe that the proposition of novel sampling heuristics that are able to maintain the original modular structure will improve the efficiency of the proposed technique both in synthetic and real networks.}

\section{Conclusion} \label{conclusao}

We have demonstrated that the walktrap and fast greedy algorithms are suitable to accurately identify communities even if many nodes of the real network were missing, which is a key issue in network theory for the many cases of incomplete information. Inspired by this robust behaviour, we devised a technique to detect the modular structure of dense networks (such as similarity networks) that is based on the application of standard methods in sampled networks. Our method provided high accuracy rates while improving the time performance in networks.

As for future work, we are planning to devise an approach to identify automatically the best sampling rate that provides optimized gain in temporal complexity, given a fixed margin of error in accuracy. An important adaptation will be developed to adapt the algorithm in networks with lower average degree. We also intend to conceive novel ways to propagate the memberships of nodes in step 5 of our method through techniques similar to those used in semi-supervised pattern recognition~\cite{teago}. Another possibility is to investigate the applicability of novel sampling techniques to further improve the accuracy and time performance. Finally, one could verify the effect of sampling in multi-resolution community analysis~\cite{newref5} and in networks with overlapping community structure~\cite{newref6}.

\section*{Acknowledgments}
The authors are grateful to S\~ao Paulo Research Foundation (FAPESP) (grant numbers 2013/06717-4 and 2014/20830-0).

\end{document}